\documentclass[twocolumn,prl,aps,superscriptaddress]{revtex4}
\usepackage{array}
\usepackage{amsmath}
\usepackage{graphicx}
\usepackage{graphicx}
\usepackage{amsmath}
\usepackage{latexsym}
\usepackage{amsfonts}
\usepackage{amssymb}
\usepackage{array}
\usepackage{epsfig}

\begin{document}

\title{Scheme for proving the bosonic commutation relation using single-photon interference}

\author{M. S. Kim}
\affiliation{School of Mathematics and Physics, The Queen's
University, Belfast BT7 1NN, United Kingdom}
\author{H. Jeong}
\affiliation{Department of Physics and Astronomy, Seoul National University, Seoul, 151-742, South Korea}
\author{A. Zavatta}
\affiliation{Istituto Nazionale di Ottica Applicata (CNR), L.go E. Fermi, 6, I-50125, Florence, Italy}
\author{V. Parigi}
\affiliation{LENS and Department of Physics, University of Firenze, 50019 Sesto Fiorentino, Florence, Italy}
\author{M. Bellini}
\affiliation{Istituto Nazionale di Ottica Applicata (CNR), L.go E. Fermi, 6, I-50125, Florence, Italy}
\affiliation{LENS and Department of Physics, University of Firenze, 50019 Sesto Fiorentino, Florence, Italy}

\date{\today}

\begin{abstract}
We propose an experiment to directly prove the commutation relation between bosonic annihilation and
creation operators, based on the recent experimental success in single-photon subtraction and addition.
We devise a single-photon interferometer to realize coherent superpositions of two sequences of photon
addition and subtraction.   Depending on the interference outcome, the commutation relation is directly
proven or a highly nonclassical state is produced.  Experimental imperfections are assessed to show that
the realization of the scheme is highly feasible.
\end{abstract}


\maketitle

The uncertainty principle, which is at a rudiment of quantum physics, is due to the non-zero commutation
relation between complementary observables.  The quantum algebra of the commutation relation plays an
important role in many of the paradoxes and applications of quantum physics.  In particular, the bosonic
commutation relation
\begin{equation}
[\hat{a},\hat{a}^\dag]=\openone \label{comm}
\end{equation}
between creation $\hat{a}^\dag$ and annihilation $\hat{a}$ operators is one of the fundamental ones,
which is directly related to the commutation relation between position and momentum observables.

On the other hand, the wave-particle duality is another important doctrine of quantum physics.  The
beauty of quantum physics lies in the fact that we can explain the seemingly contradictory riddle of
duality using one theory. Single-photon interference is one of the examples due to the duality of nature
and accurately interpreted by quantum physics.   The duality and uncertainty principle have been studied
since the birth of quantum physics and there have been many experimental evidences which confirm quantum
mechanical predictions.  However, as far as we are aware, there has not been a direct proof of the
bosonic commutation relation, which we are proposing in this paper based on single-photon interference.

We have recently witnessed experimental successes  in photon-level operations to
subtract~\cite{grangier1} and add~\cite{bellini1} a single photon in a light field.  These prove to be
important as they are essential building blocks for quantum-state engineering and provide a tool to
experimentally show the foundations of quantum mechanics~\cite{boyd}. It is remarkable that adding a
definite number of photons any classical field can be made into a nonclassical one as its statistics in
phase space cannot be described by a classical theory~\cite{mandel}.  For the recent interest of quantum
entanglement, it has been experimentally shown~\cite{our} that entanglement can be enhanced by
subtracting a photon from one of the two modes of a two-mode squeezed state.

Before providing details of our proposal to directly prove the bosonic commutation relation, we briefly
describe the single-photon-level operations involved.  We then devise a single-photon interferometer
using the wave-particle duality to interfere two subtraction processes and show the commutation relation
between subtraction and addition processes.  Heralded by the interference outcome, we can also produce a
highly nonclassical state.

Let us consider a photon-subtraction~\cite{grangier1} scheme recently realized~\cite{grangier2,bellini2}.
A photon is subtracted from input state
$|\psi\rangle=\sum_{n=0}^{\infty}C(n)|n\rangle$ by splitting out a single photon using a lossless beam
splitter of high transmittivity, $t$ (low reflectivity $r$) and a photodetector.  With use of the
standard form of a beam splitter operator $\hat{B}(t)$ ~\cite{kim1}, we find the beam splitter output
for the input state $|\psi\rangle$ in mode $a$, as
\begin{equation}
\hat{B}(t)|\psi\rangle|0\rangle=\sum_{n=0}^\infty C(n)
\sum_{k=0}^n \left(\begin{array}{c} n \cr k \cr
\end{array}\right)^{1\over 2}r^kt^{n-k}|n-k\rangle |k\rangle
\label{bs-photon-1}
\end{equation}
assuming nothing (vacuum state $|0\rangle$) has been injected into the other input port of mode $b$. 
The binomial coefficient has been denoted by $\tiny{\left(\begin{array}{c} n \cr k \cr \end{array}\right)}$.
With $T\approx 1$, if one particle is found at the output
$b_{out}$, the conditional probability of having $n-1$ particles at output $a_{out}$ is approximated as
follows:
\begin{equation}
P_{sub}(n-1)={\cal N}_s\frac{n!}{(n-1)!}T^{n-1} P_0(n)\approx {\cal N}_snP_0(n)\label{bs-2}
\end{equation}
taking $T=t^2$ and $P_0(n)=|C(n)|^2$.  Throughout the paper, ${\cal N}$ with a subscript denotes 
a respective normalization factor.  The proportionality on $n$  in Eq.~(\ref{bs-2}) reflects the 
coefficient $\sqrt{n}$ which emerges when the annihilation operator $\hat{a}$ is applied to a Fock state
$|n\rangle$~\cite{more}.
 As far as the photon number statistics is concerned, the subtraction scheme can
be understood by treating the photons as conventional particles and assuming a beam splitter as a device
which randomly chooses incoming particles to change their directions with the probability $1-T$.

Recently, a very neat scheme to add a photon~\cite{bellini1} has been realized using a parametric
downconverter which produces twin photons to modes $a$ and $b$ and is described by
$\hat{S}(s)=\exp(-s\hat{a}^\dag\hat{b}^\dag+s\hat{b}\hat{a})$ with the coupling parameter
$s$~\cite{barnett}.  For the input state $|\psi\rangle$ in mode $a$ and the vacuum in the ancilla mode
$b$, the output state is
 \begin{equation}
\hat{S}(s)|\psi\rangle|0\rangle =\sum_{n=0}^\infty \frac{C(n)}{\mu^{n+1}}\sum_{k=0}^\infty
(-\lambda)^k\sqrt{\frac{(n+k)!}{k!^2}}|n+k\rangle |k\rangle, \label{am-photon-1}
\end{equation}
where $\mu=\cosh s$, $\nu=\sinh s$ and $\lambda=\nu/\mu$.  Once a photon is detected in output mode $b$,
we find that the state in Eq.~(\ref{am-photon-1}) brings about the conditional probability
$P_{add}(n+1)\approx {\cal N}_a(n+1)P_0(n), \label{amp-2}$
assuming $\mu\approx 1$.  Here, the factor $n+1$ is the realization of $\sqrt{n+1}$ as the creation
operator $\hat{a}^\dag$ acts on $|n\rangle$~\cite{more}.

In \cite{bellini2}, the authors compare two sequences of photon addition and subtraction and show the
quantum nature of the operations through the photon number distribution and the phase-space statistics.
However, it fails to show the exact commutation relation other than the difference between the two
sequences.  In the experiment~\cite{bellini2} for a thermal input field, it has been found that the mean
photon number after the sequence of photon subtraction then addition ($\hat{a}^\dag\hat{a}$) is larger
by one photon than that after the sequence of photon addition then subtraction ($\hat{a}\hat{a}^\dag$).
At the first glance, this is odd because the commutation relation (\ref{comm}) seems to advocate the
other way round.  However, the mean values are obtained after the normalization of the density operators
to show only the statistical averages, thus they cannot reveal the commutation relation directly.  This
is why we need to carefully design a new setup for its direct proof.

{\it Direct proof of commutation relation}.- The density operator of a field obtained by adding a photon
after subtracting one is $ \hat{\rho}_1={\cal N}_1\hat{a}^\dag\hat{a}\hat{\rho}_0\hat{a}^\dag\hat{a} $
where $\hat{\rho}_0$ is the density operator for the initial field.  On the other hand, by subtracting a
photon after adding one the density operator becomes $ \hat{\rho}_2={\cal
N}_2\hat{a}\hat{a}^\dag\hat{\rho}_0\hat{a}\hat{a}^\dag. $ Once these two experiments are {\it
separately} performed, it is not possible to show the commutation relation directly from the
experimental data.  For example, the photon number distributions will tell us about the difference
between $\hat{a}^\dag\hat{a}\hat{\rho}_0\hat{a}^\dag\hat{a}$ and
$\hat{a}\hat{a}^\dag\hat{\rho}_0\hat{a}\hat{a}^\dag$ rather than
$(\hat{a}\hat{a}^\dag-\hat{a}^\dag\hat{a})\hat{\rho}_0 (\hat{a}\hat{a}^\dag-\hat{a}^\dag\hat{a})$.  It
is then clear that we should have a coherent superposition of two sequences of operations through their
interference, to directly show the commutation relation.   This is an interesting remark because the
concept of interference is associated to the wave nature.

Let us consider the setup in Fig.~1.  The  beam splitters, $\rm BS1$ and $\rm BS2$, with the same high
transmittivity subtract photons from the input field.  A parametric downconverter produces twin photons
into two different modes.  A photon counting at PD0 heralds that a photon has been added to the input
field which passed through the downconverter. The two reflected fields at $\rm BS1$ and $\rm BS2$
interfere at the 50:50 beam splitter $\rm BS3$ to erase information about their paths.   Had there not
been $\rm BS3$, one photon detected in mode $b$ but not in mode $c$ indicates that one photon has been
subtracted before the photon addition process.  On the contrary, if one photon is detected in mode $c$,
we know that a photon subtraction has been performed after the photon addition. However, as the
photodetectors, $\rm PD1$ and $\rm PD2$, are placed after $\rm BS3$, by having one photon registered in
either of the photodetectors, there is no way to find out if the photon was subtracted before or after
the photon addition process. Thus detecting one photon is to herald a superposition of two possible
sequences $\hat{a}^\dag\hat{a}$ and $\hat{a}\hat{a}^\dag$.

A beam splitter is a unitary operator and its action is described by the following input-output relation
for two modes $b$ and $c$:
\begin{equation}
\hat{B}(t)\left(\begin{array}{c} \hat{b} \cr \hat{c} \cr
\end{array}\right)_{in}\hat{B}^\dag(t)=\left(\begin{array}{c}t\hat{b}+r\hat{c}
\cr  t\hat{c}-r\hat{b}\end{array}\right)_{out}.
\label{bs-uni}
\end{equation}
For a 50:50 beam splitter like $\rm BS3$, the reflectivity $r={1\over\sqrt{2}}$.  The two different
signs of $\pm$ in the right-hand side of Eq.~(\ref{bs-uni}) ensure the unitarity of the beam splitter
operator and play a key role in the bunching of two photons for the Hong-Ou-Mandel
interferometer~\cite{hong}.  With use of Eq.~(\ref{bs-uni}), we find that if a photon is detected at
$\rm PD1$ but not at $\rm PD2$, the operation by the whole set up is
$\hat{a}^\dag\hat{a}+\hat{a}\hat{a}^\dag$, assuming a photon is added
between $\rm BS1$ and $\rm BS2$.  Note that the overall operation is the constructive interference 
of two operations both of which transform an initial state into nonclassical ones. On the
other hand, if a photon is detected at $\rm PD2$ instead of $\rm PD1$, the operation is
$\hat{a}\hat{a}^\dag-\hat{a}^\dag\hat{a}=\openone$, which means that the conditional output field should
be identical to the input field.  {\it This is the direct proof of the bosonic commutation relation}~\cite{more2}.

\begin{figure}[t]
\begin{center}
\centerline{\includegraphics[width=3.4in]{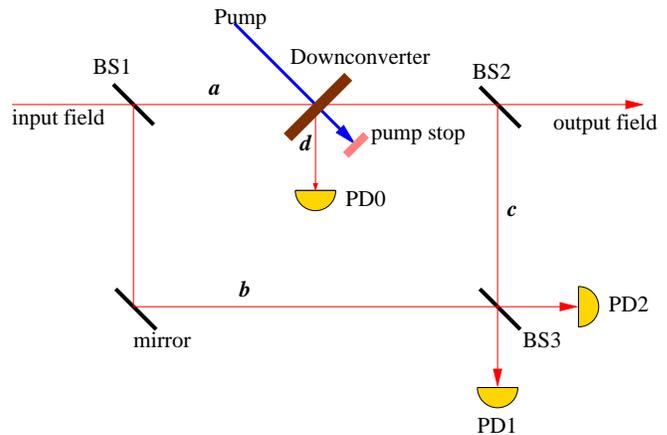}}
\end{center}
\caption{Experimental setup. $\rm BS1$ and $\rm BS2$ are beam
splitters of high transmittivity.  A photon is added by a parametric downconverter between $\rm BS1$ and
$\rm BS2$.  A 50:50 beam splitter ($\rm BS3$) superposes the reflected fields from $\rm BS1$
and $\rm BS2$.  The output field is selected, conditioned on registering a photon at only one of the two
photodetectors $\rm PD1$ and $\rm PD2$. $a$, $b$, $c$ and $d$ label the field modes. }
\end{figure}

We now show how our scheme works by following each step carefully.
A general pure input state can be written  as
$|\phi_0\rangle$. 
Once the pure state case is clear, the extension to a mixed state input is straightforward.
We assume nothing is injected to the unused input ports of $\rm BS1$ and $\rm BS2$.  In the following argument, normalization is not included for simplicity.  According to Ref.~\cite{more}, the action of $\rm BS1$ under the condition $t^{\hat{a}^\dag\hat{a}}\approx \openone$, 
\begin{equation}
\hat{B}_{ab}|\phi_0, 0\rangle_{ab}\approx(1+{r \over t}\hat{a}\hat{b}^\dag)|\phi_0, 0\rangle_{ab}.
\label{supple-1}
\end{equation}
The subscripts $a,b,\cdots$ denote modes in Fig.~1. With $\mu\approx1$ and $\nu\ll 1$, we approximate
$\hat{S}|00\rangle \approx (1-\lambda\hat{a}^\dag\hat{d}^\dag)|00\rangle$
again without normalization.  We measure a photon at photodetector $\rm PD0$, then the conditioned state is described by 
\begin{eqnarray}
&&_d\langle 1|\hat{S}_{ad}(1+{r\over t}\hat{a}\hat{b}^\dag)|\phi_0,0, 0\rangle_{abd}
\nonumber\\
&&~~~~~~~\approx (-\lambda\hat{a}^\dag-{r\over t}\lambda\mu\hat{a}\hat{a}^\dag\hat{b}^\dag+{r\over t}\nu\hat{b}^\dag) |\phi_0,0\rangle_{ab},
\label{supple-2}
\end{eqnarray}
where the unitary operation $\hat{S}\hat{a}\hat{S}^\dag=\mu\hat{a}+\nu\hat{d}^\dag$ has also been used.  Passing through $\rm BS2$ of the same high transmittivity as $\rm BS1$, the state of the field modes $a$, $b$ and $c$ becomes
\begin{equation}
-\lambda r\left[(1+\hat{a}^\dag\hat{a})\hat{c}^\dag+(t\mu\hat{a}\hat{a}^\dag-{\nu\over t\lambda})\hat{b}^\dag\right]|\phi_0,0,0\rangle_{abc}.
\label{supple-3}
\end{equation}
As we impose a condition $t\mu\approx {\nu\over t\lambda}\approx 1$, which is well satisfied for the proposed experimental scheme,
the output field after $\rm BS2$ and $\rm BS3$ is approximated to $-\lambda r(\hat{a}\hat{a}^\dag\hat{c}^\dag+\hat{a}^\dag\hat{a}\hat{b}^\dag)|\phi_0,0,0\rangle_{abc}$. Now,  by the final 50:50 beam splitter {\rm BS3}, the field becomes
\begin{equation}
{\lambda r \over \sqrt{2}}[(\hat{a}\hat{a}^\dag-\hat{a}^\dag\hat{a})\hat{b}^\dag-(\hat{a}\hat{a}^\dag+\hat{a}^\dag\hat{a})\hat{c}^\dag]|\phi_0,0,0\rangle_{abc}
\label{supple-4}
\end{equation}
which shows that one photon detected in mode $b$ by $\rm PD2$ and none in mode $c$ should result in a unit operation $\openone$ while in mode $c$ by $\rm PD2$ (none in mode $b$) the operation is $\hat{a}^\dag\hat{a}+\hat{a}\hat{a}^\dag$.

{\it Experimental feasibility}.- 
There are some details we have to consider for experimental feasibility of our scheme. 
One problem
comes from the fact that there is no photon-level detector available.  Thus we have to
replace the photon number resolving detectors in our theory
with on-off type detectors realized by avalanche photodiodes, which discern
there being photons from no photons with high efficiency. The `on' event is represented by
$\openone -|0\rangle\langle 0|$ and `off' by $|0\rangle\langle 0|$.  

We exemplify the effect of the realistic experimental condition for the coherent input state,
$|\alpha\rangle$, which is at the boundary between quantum and classical worlds.
Then the output state conditioned on the 
photodection at the avalanche photodiode {\rm PD2} but not at {\rm PD1} is found to be 
$\approx|t\alpha\rangle$ (In fact the resultant state is mixed but other terms 
are negligible).
In order to assess the closeness between the input state $|\phi_0\rangle$
and the conditional output state of density operator $\hat{\rho}_{out}$,
the fidelity defined as $F=\langle\phi_0|\hat{\rho}_{out}|\phi_0\rangle$ is utilized.
Taking realistic experimental values of $T=0.99$ and $s=0.1$, we find that
the fidelity $F\approx\mbox{e}^{-(1-t)^2|\alpha|^2}$.
The fidelity is as high as $F>99.99\%$ for $|\alpha|^2\lesssim 2$.
 Thus, the unit operation ($\openone$) which shows the bosonic commutation relation can be proven efficiently. 
 
 \begin{figure}[t]
\begin{center}
\vspace{0.5cm} {\bf (a)}\hskip4cm{\bf (b)}
\centerline{\psfig{figure=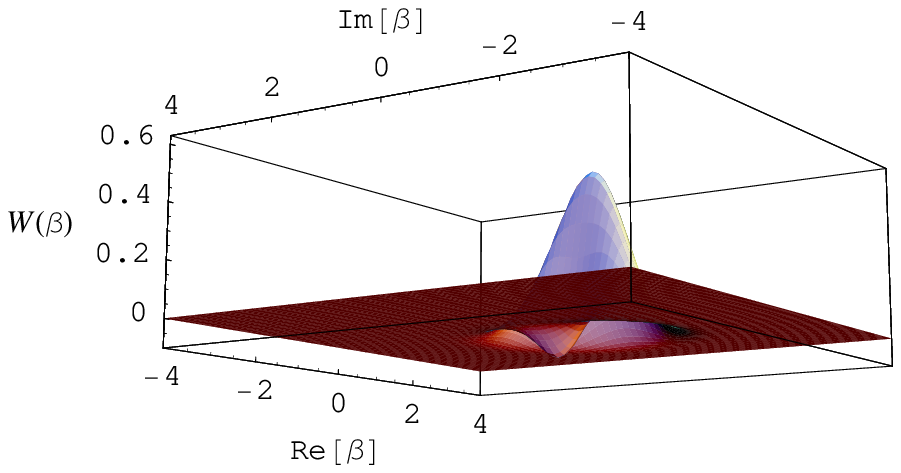,width=4cm,height=2.95cm}
\psfig{figure=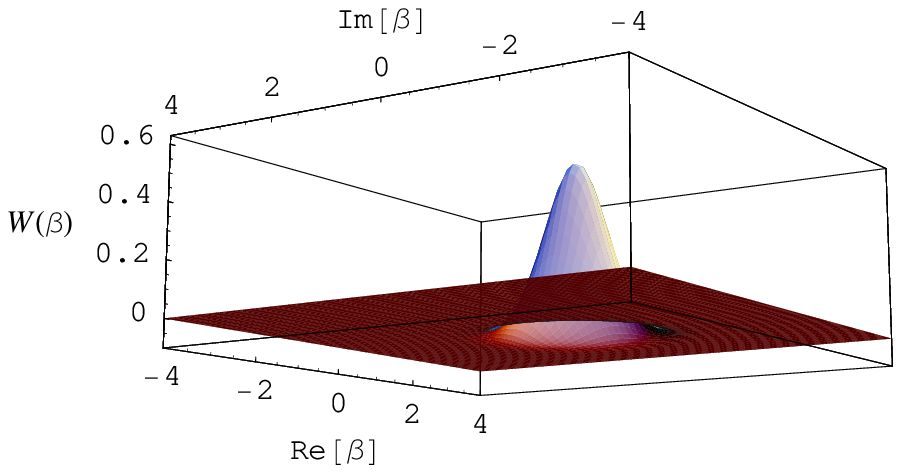,width=4cm,height=2.95cm}}
\end{center}
\caption{(Color online): Wigner function $W(\beta)$ for the output state conditioned on photodetection
only at $\rm PD1$ {\bf (a)} or only at $\rm PD2$ {\bf (b)}
when the initial state is a coherent state of amplitude $\alpha=1$. $T=0.99$ and
$\mu=1.005$ ($s=0.1$).} \label{fig:plot2}
\end{figure}
 
It is useful to represent a quantum state in phase space by quasi-probability functions as they
visualize the quantum state and can be used to show its nonclassical nature.
In particular, we utilize the Wigner function $W(\beta)$, where the real and imaginary parts
of $\beta$ are two conjugate variables (see \cite{barnett} for its definition).
It is well known that the Wigner function may show negative values 
reflecting the nonclassical nature of a given state.
Now, the Wigner function
 for the output state conditioned on
the photodetection only at the avalanche photodiode $\rm PD1$ ($\rm PD2$) for the initial coherent state of
 $\alpha=1$, is shown in
Fig.~2(a) (Fig.~2(b)).  We can clearly see the negativity around the origin of the phase space in (a) which 
is contrasted to the positive Gaussian Wigner function in (b).

The inefficiency of avalanche photodetectors  was not a
problem in separate photon addition and subtraction experiments because it only lowers the success
probability. However, in our proposal, another important fact is to make sure that one output port of
$\rm BS3$ is empty while the other registers a photon. If the detection efficiency is low, only one of
the detectors may click while the other is silent even though there are photons at both modes $b$ and
$c$ after $\rm BS3$. However, we stress that this ``failure" probability is very low regardless of the
detection inefficiency. It is straightforward to obtain the probability for both the modes $b$ and $c$
having photon(s) before the final detection as 
${\cal P}_{bc}={\rm Tr}[\hat{\rho}_{out}\openone_a \otimes(\openone-|0\rangle
\langle0|)_b\otimes (\openone-|0\rangle \langle0|)_c]$. Then, it can be conditioned on the probability
of the detection at $\rm PD1$, ${\cal P}_{b}={\rm Tr}[\hat{\rho}_{out}\openone_a \otimes(\openone-|0\rangle
\langle0|)_b\otimes \openone_c]$, so that the conditional probabilities ${\cal P}_{bc|b}={\cal P}_{bc}/{\cal P}_b$
(and ${\cal P}_{bc|c}$ in the same manner) can be obtained. 
These conditional probabilities are only ${\cal P}_{bc|b}\approx0.2\%$
(${\cal P}_{bc|b}\approx 0.2\%$) and
${\cal P}_{bc|c}\approx 2\%$ (${\cal P}_{bc|c}\approx 1\%$)  for $T=0.99$, $s=0.1$, with the
initial coherent state of amplitude $\alpha=1$ ($\alpha=0.6$).
This means that our scheme is robust against the detection inefficiency.
For example, suppose that the detection efficiency is 45\% for both PD1 and PD2
and we look at the case of photodetection only at PD2 to prove the commutation relation.
A simple analysis based on the aforementioned values immediately
leads to the conclusion that the degradation of the fidelity is
less than 1.1\% (0.55\%) for $\alpha=1$ ($\alpha=0.6$).

The conditionally-prepared state in mode $a$
can be completely characterized by means of 
homodyne detection. This
allows one to reconstruct its Wigner function $W(\beta)$ in phase space. 
In an experiment involving homodyne detection, a thermal state, which is 
a bosonic state in thermal equilibrium at a given temperature~\cite{barnett}
 and can also be implemented by phase and amplitude
randomization of a coherent field, may be handier to use as an input,
because it does not require precise phase control of the local oscillator.
Consider that the initial field is a thermal field of mean photon number $\bar{n}$.  Figs.~3 show the
Wigner functions for the output field conditioned on photodetection at $\rm PD1$ in (a) and $\rm PD2$ in (b). 
While the Wigner function in (b) shows a Gaussian profile as for the
initial thermal field, that in (a) shows a negative dip at the origin, manifesting nonclassicality.
The levels of homodyne detection efficiency reached in current experiments guarantee that these effects should
be clearly visible in a realistic situation \cite{grangier2,bellini2}. The dark count rate of photodetectors 
could generally be neglected
in photon subtraction and addition experiments \cite{grangier1,bellini1,
our,grangier2,bellini2}. 

\begin{figure}[t]
\vspace{0.5cm} {\bf (a)}\hskip4cm{\bf (b)}
\centerline{\psfig{figure=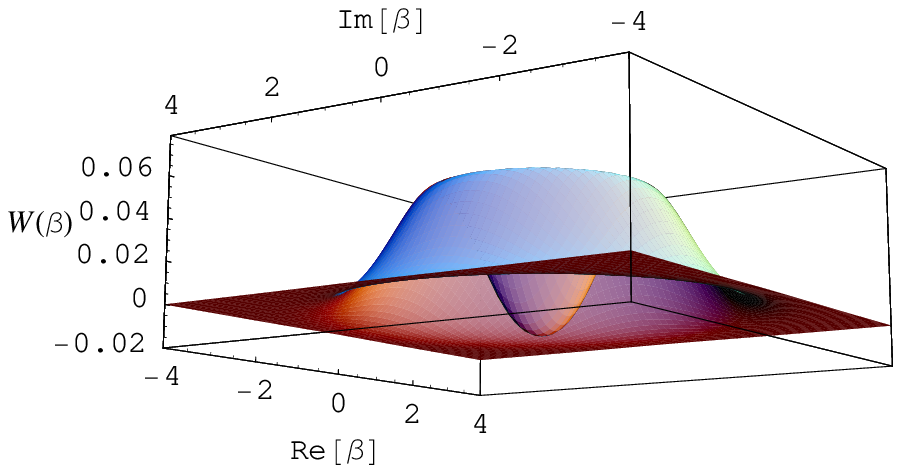,width=4cm,height=2.95cm}\psfig{figure=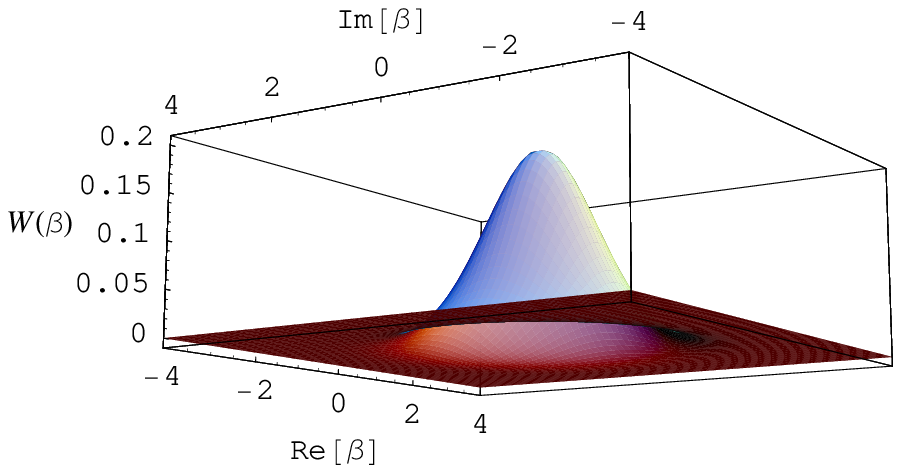,width=4cm,height=2.95cm}}
\caption{(Color online) Wigner functions $W(\beta)$ for the output field of mode $a$ for photodetection
only at $\rm PD1$ {\bf (a)} or only at $\rm PD2$ {\bf (b)}. The average photon number of the initial
thermal field is $\bar{n}=1$. $T=0.99$ and $\mu=1.005$.} \label{fig:plot3}
\end{figure}

{\it Remarks}.- We have devised a single-photon interferometer based on photon addition and subtraction
techniques, realized in numerous labs worldwide.  Our interferometer will enable the first direct test
of the bosonic commutation relation as it superposes two different sequences of operations. 
 Heralded by the interference outcome, we can also produce a
nonclassical state which may be very different from the initial state.  The assessment of experimental
inefficiencies suggest that the scheme can be readily implemented with high feasibility.

{\it Acknowledgements.}- MSK thanks Prof. Englert for showing him complementarity in an optical setup and
Dr. Dunningham for discussions.  This work was supported by Ente Cassa di Risparmio di
Firenze, by MIUR-PRIN, CNR-RSTL, EPSRC, QIP IRC, and KOSEF through the Center for Subwavelength Optics.


\begin{thebibliography}{99}
\bibitem{more} It can be shown that $\hat{B}|\psi\rangle_a|0\rangle_b=\mbox{e}^{{r\over t}\hat{a}\hat{b}^\dag}t^{\hat{a}^\dag\hat{a}}|\psi\rangle_a|0\rangle_b$.  Under our condition $t\approx 1$,  and $|\psi\rangle$ not to have significant contribution from highly excited states, $\hat{B}|\psi\rangle_a|0\rangle_b\approx(1-{r\over t}\hat{a}\hat{b}^\dag)|\psi\rangle_a|0\rangle_b$, which becomes an annihilation operation when a single photon is measured in mode $b$: $_b\langle
1|\hat{B}|\psi\rangle_a|0\rangle_b\propto\hat{a}|\psi\rangle_a$.  Analogously, we can show that  $_b\langle 1|\hat{S}|\psi\rangle_a|0\rangle_b\propto\hat{a}^\dag |\psi\rangle_a$ under $\mu\approx 1$ and no high excitations in the initial field.(See \cite{kim} for details.)
\bibitem{more2} Because of the normalization involved, the proof of the final state being equal to the initial state is enough only to say that the commutation relation bears a constant value. In order to find this constant value, the anti-commutator operation heralded by PD2 detection can, for example, be utilized.
\end{thebibliography}
\end{document}